\pdfoutput=1

\documentclass[journal]{IEEEtran}
\ifCLASSINFOpdf
\else
\fi
\usepackage{array}
\newcolumntype{P}[1]{>{\centering\arraybackslash}p{#1}}
\usepackage{algorithm}
\usepackage{algorithmicx}
\usepackage{algpseudocode}
\usepackage{tabu}
\usepackage{amssymb}
\usepackage{mathrsfs}
\usepackage{amsmath,cite}
\usepackage{chemarrow}
\usepackage{hyperref}
\usepackage{url}
\usepackage{color}
\usepackage{graphicx}          % Include this line if your
\usepackage{cite} 
\usepackage{epstopdf}
\newtheorem{rem}{Remark}
\newtheorem{prop}{Proposition}
\newtheorem{lem}{Lemma}
\begin{document}
\title{Sparse Bayesian Inference of Multivariable ARX Networks}

\author{Junyang Jin, Ye Yuan, Alex Webb and Jorge Gon\c{c}alves% <-this % stops a space 
\thanks{Junyang~Jin and Alex Webb are with Circadian Signal Transduction Group, Department of Plant Sciences, University of Cambridge. Ye~Yuan is with School of Automation, Huazhong University of Science and Technology. Jorge~Gon\c{c}alves is with the Department of Engineering, University of Cambridge and the Luxembourg Centre for Systems Biomedicine. (Corresponding author: Ye Yuan).}}% <-this % stops a space
\maketitle

\begin{abstract}
Increasing attention has recently been given to the inference of sparse networks. In biology, for example, most molecules only bind to a small number of other molecules, leading to sparse molecular interaction networks. 
To achieve sparseness, a common approach consists of applying weighted penalties to the number of links between nodes in the network and the complexity of the dynamics of existing links. The selection of proper weights, however, is non-trivial. 
Alternatively, this paper proposes a novel data-driven method, called GESBL, that is able to penalise both network sparsity and model complexity without any tuning. GESBL combines Sparse Bayesian Learning (SBL) and Group Sparse Bayesian Learning (GSBL) to introduce penalties for complexity, both in terms of element (system order of nonzero connections) and group sparsity (network topology). 
The paper considers a class of sparse linear time-invariant networks where the dynamics are represented by multivariable ARX models. 
Data generated from sparse random ARX networks and synthetic gene regulatory networks indicate that our method, on average, considerably outperforms existing state-of-the-art methods. The proposed method can be applied to a wide range of fields, from systems biology applications in signalling and genetic regulatory networks to power systems.\footnote{Part of this work has been presented in~\cite{Jin}. This paper, considerably improves~\cite{Jin} with two additional algorithms and new simulations that demonstrate the advantages of GESBL.}
\end{abstract}

\begin{IEEEkeywords}
System Identification, Sparse Bayesian Learning, Polynomial Model, Network Inference, Sparse Networks, Systems Biology.
\end{IEEEkeywords}

% \fi
\IEEEpeerreviewmaketitle

\section{Introduction}
When designing feedback controllers, it is typically sufficient to learn the input-output dynamics of the system, independently of its internal complexity. Hence, most of the work on system identification focuses on modeling input-output dynamics without exploring the internal topology. Yet, in many applications, information about the network topology is critical. For example, we may require the internal topology and dynamics of a system to understand its mechanisms of action or to locate the source of faults. Examples range from biomedicine to autonomous vehicles, power and communication networks. 

Sparsity is an inherent property of many important networks. In biology, most molecules bind with a small number of other molecules. In autonomous vehicles, communication can be constrained to neighbours to minimise energy consumption. Elements of power and communication networks are typically connected to a small number of other elements. Hence, sparsity can be used as a constraint to model networks and to compensate for potentially low number of noisy samples. 

Standard system identification methods, such as the prediction error method (PEM) or Maximum-likelihood (ML), are applicable to a large family of black-box models, including ARX, ARMAX and Box-Jenkins~\cite{sys}. However, these methods alone fail to capture the sparsity feature of networks. For noisy MIMO systems, where there is no prior knowledge of the topology, PEM generates full transfer matrices even if the ground truths are sparse~\cite{nonp,yuan2011robust}. Hence, methods must penalise model complexity to favour sparsity.

Maximum a posteriori methods (MAP Type I method) include Least Absolute Shrinkage and Selection Operator (LASSO), Tikhonov regularisation, FOcal Underdetermined System Solver (FOCUSS) and Sparse Group LASSO (SGL)~\cite{thesis,SGL}. All these methods penalise model complexity. For example, a LASSO algorithm has been used to infer the topology of linear MIMO systems from steady-state data~\cite{sysc}. Similar work has inferred sparse multivariable ARX models with known polynomial order using a greedy algorithm, Block Orthogonal Matching Pursuit (BOMP), which focuses more on identifying the network topology~\cite{bomp}.  Whilst these approaches effectively reduce over-fitting, the weighting variable, which controls the trade-off between data-fitting and model complexity (sparsity), must either be chosen {\em a priori} or evaluated independently, using methods such as cross-validation. Unfortunately, this increases the computational burden and causes information waste. 

Some alternative methods do not require a tuning variable. These include Sparse Bayesian Learning (SBL; a type II method) and kernel methods. Both methods apply the technique of Bayesian approximation: the difference between them lies in the use of contrasting kernel functions. In contrast to MAP methods, SBL, a well-known method in machine learning, applies inseparable priors~\cite{group1,tipp,gsbl,prior1}. It has been applied to identify nonlinear systems by selecting nonlinear functions from a predefined dictionary~\cite{statewei}. The nonlinear model structure is captured either by element SBL or by Group Sparse Bayesian Learning (GSBL), depending on the type of data available. However, full state measurements are required. The kernel method, in contrast, is a non-parametric approach, introduced to estimate impulse responses of SISO systems~\cite{nonp2}. It has been combined with empirical Bayes to identify discrete-time linear systems (e.g. ARMAX)~\cite{nonp,nonp1,nonp3}. 

The above methods focus on either element sparsity (system order) or group sparsity (network topology). However, the identification of ARX networks simultaneously requires both kinds of sparsity, which is the main focus of this paper. The goal are two-fold: (a) to infer the network topology and (b) to estimate model parameters including polynomial orders. 

While MAP approaches normally demand extra efforts in estimating tuning variables, our methodology is tuning-free. The identification problem is formulated as a linear regression, where the target vector is both group and element sparse. This is achieved by combining SBL with GSBL to simultaneously achieve both kinds of sparsity. Simulations on randomly generated sparse networks show that our method (denoted by GESBL) outperforms SBL, GSBL and the kernel method. The evident improvement in detecting sparse topology is reflected by simulations of ring structure networks. Further simulations of a realistic biological network model show that GESBL outperforms the state-of-the-art method, iCheMA.

The paper is organized as follows. Section \ref{sec:Model} introduces the polynomial model and discusses its identifiability. Section \ref{sec:Reconstruction} formulates the network reconstruction problem.  Section~\ref{sec:Bayesian} promotes a sparse prior and discusses three algorithms to solve the resultant optimisation problem. Section~\ref{ndsf} considers an extended nonlinear polynomial model. Section~\ref{sec:Simulation} compares the method with other approaches via Monte Carlo simulation. Finally, Section~\ref{sec:Conclusion} concludes and discusses further development in this field.

\emph{Notation}: The notation in this paper is standard. $I$ denotes the identity matrix. For $L\in{R}^{n\times n}$, $diag\{L\}$ denotes a vector which consists of diagonal elements of matrix $L$ and $[L]_{ij}$ presents the $ij$th entry. $blkdiag\{L_1,...,L_n\}$ is a block diagonal matrix. $trace\{L\}$ denotes the trace of the matrix. $L\succeq0$ means $L$ is positive semi-definite. $ \|w\|^{2}_{L^{-1}}$ represents $w^TL^{-1}w$. For $l\in{R}^{n}$, $diag\{l\}$ denotes a diagonal matrix whose diagonal elements come from vector $l$. $[l]_{ij}$ denotes the $j$th element of the $i$th group of $l$. $l\geq0$ means each element of the vector is non-negative. $v=l^{k}$ is also a vector where $v_i=(l_i)^k$.  $\text{vec}\{x_1,..,x_n\}=\left[x_1,...,x_n\right]'$ means to vectorise elements $\{x_1,..,x_n\}$. A vector $y(t_1 : t_2)$ denotes a row vector $\left[\begin{array}{cccc}y(t_1)&y(t_1 +1)&\cdots&y(t_2)\end{array}\right]$. 

\section{MODEL SPECIFICATION}\label{sec:Model}
The sparse network of $p$ nodes and $m$ inputs is described by a parametrised multivariable ARX model, $\mathcal{M^*}(w^*)$:
$A(z^{-1};w^*)Y(t) =  B(z^{-1};w^*)U(t) + E(t),$
where
\begin{equation}
\begin{aligned}
A(z^{-1};w^*) &=  I + \hat{A}_1z^{-1}+...+\hat{A}_{n_a^*}z^{-n_a^*},\\
B(z^{-1};w^*) &=   \hat{B}_1z^{-1}+...+\hat{B}_{n_b^*}z^{-n_b^*}.
\label{pARX}
\end{aligned}
\end{equation}
$z^{-1}$ is the time shift operator. $Y(t)\in{R}^{p}$ represents the nodes of the network, $U(t)\in{R}^{m}$ denotes the input, and $E(t)\in{R}^{p}$ is i.i.d. white Gaussian noise. $\hat{A}_i\in R^{p\times p}$ and $\hat{B}_i\in R^{p\times m}$ are matrices. $w^*$ contains all the model parameters. $A(z^{-1};w^*)$ is a polynomial matrix showing the connectivity of each node to the others including self-loops. Similarly, $B(z^{-1};w^*)$ is a polynomial matrix relating the input to the nodes. The Boolean structure of the network is reflected by the nonzero elements in $A(z^{-1};w^*)$ and $B(z^{-1};w^*)$ whereas the system dynamics are dominated by the Input-Output map of the model: 
\begin{equation}
\begin{aligned}
Y(t) =  G_u(z^{-1};w^*)U(t) + G_e(z^{-1};w^*)E(t),
\label{inout}
\end{aligned}
\end{equation}
where $G^*(z^{-1};w^*)$ denotes the transfer matrix of the model, $\mathcal{M^*}(w^*)$:
\begin{equation}
\begin{aligned}
G_u(z^{-1};w^*) &= A^{-1}(z^{-1};w^*)B(z^{-1};w^*), \\
G_e(z^{-1};w^*) &=   A^{-1}(z^{-1};w^*),\\
G^*(z^{-1};w^*) &=   \left[G_u(z^{-1};w^*)\ \ G_e(z^{-1};w^*)\right].
\end{aligned}
\end{equation}
 It is important to ensure that ARX models are identifiable. It is shown in~\cite{sys} that multivariable ARX models are strictly globally identifiable, provided the order of the system is known. However, in practice the order of the system is usually unknown and, hence, it must also be estimated. For that, it is common to turn to information criteria methods, such as AIC and BIC. To avoid combinatorial search of the system order, SBL is applied in our framework.
%  Other approaches apply a model set, $\mathcal{M}(w)$ that contains the ground truth model ($\exists w^\star:G(z^{-1};w^\star)=G^*(z^{-1};w^*)$) but is more complex than necessary and then penalize the model order in the designed optimisation problem. Normally, this parametrized model, $\mathcal{M}(w^\star)$ of higher order is unidentifiable due to the possible cancellation of common factors. 

For a fixed system order, $\mathcal{M^*}(w^*)$ can be represented by a higher order counterpart, $\mathcal{M}(\bar{w})$, with the coefficients associated with the excessive polynomial terms equal to $0$. Hence, the Input-Output map is retained, i.e. $G(z^{-1};\bar{w})=G^*(z^{-1};w^*)$, and $\mathcal{M}(w)$ is globally identifiable at $\bar{w}$. Moreover, with the addition of many zero parameters, the solution $\bar{w}$ is sparse, i.e. it contains many zeros. 
Thus, the basic objective is to identify the model $\mathcal{M}(\bar{w})$. From the estimation result $\bar{w}$, the ground truth polynomial order of $\mathcal{M^*}(w^*)$ is estimated as the order of the last nonzero polynomial term of $\mathcal{M}(\bar{w})$, and the model parameter $w^*$ is estimated by the nonzero elements of $\bar{w}$. 

%\IEEEpubidadjcol

\section{FORMULATION OF THE RECONSTRUCTION PROBLEM}\label{sec:Reconstruction}
Consider an ARX model $\mathcal{M}(w)$ where each node is parametrized identically. The system order $k$ is set sufficiently large (based on prior knowledge or intuition about the system). For node $i$,
\begin{equation}
\begin{aligned}
&y_{i}(t)=-[A(z^{-1})]_{i1}y_{1}(t)-\ldots +\{1-[A(z^{-1})]_{ii}\}y_{i}(t)\\
&+[B(z^{-1})]_{i1}u_{1}(t)+...+[B(z^{-1})]_{im}u_{m}(t)+e_{i}(t)
\label{armax}
\end{aligned}
\end{equation}
where $y_r(t)$ denotes the $r$th node, $u_r(t)$ the $r$th input and $e_i(t)$ i.i.d. Gaussian noise and
\begin{equation}
\begin{aligned}
\left[A(z^{-1})\right]_{ii}&=a^{ii}_{1}z^{-k}+a^{ii}_{2}z^{-k+1}+\ldots +a^{ii}_{k}z^{-1}+1,\\
\left[A(z^{-1})\right]_{ij}&=a^{ij}_{1}z^{-k}+\ldots +a^{ij}_{(k-1)}z^{-2}+a^{ij}_{k}z^{-1},\\
\left[B(z^{-1})\right]_{ij}&=b^{ij}_{1}z^{-k}+\ldots +b^{ij}_{(k-1)}z^{-2}+b^{ij}_{k}z^{-1},\\
k &\geq max\{n^*_a,n^*_b\},
\end{aligned}
\end{equation}
where $a$ and $b$ denote coefficients of polynomial terms. Superscript $ij$ represents the coefficient of  the $ij$th entry of a matrix and subscript is the index. To simplify notation, hereafter $w$ is a vector that only includes parameters of subsystem~\eqref{armax} of the $i$th node.
%\begin{equation}
%\begin{aligned}
%Y = \left[\begin{array}{c} y_1(t)\\  \vdots \\  y_{p}(t) \end{array}\right],~
%U = \left[\begin{array}{c} u_1(t)\\  \vdots \\  u_{m}(t) \end{array}\right],~
%E = \left[\begin{array}{c} e_1(t)\\  \vdots \\  e_{p}(t) \end{array}\right] \\
%\begin{array}{cc}
%Q = [Q_{ij}(z^{-1})], &Q_{ii} = 0,\\
%P = [P_{ij}(z^{-1})],&H = diag(H_{ii}(z^{-1})).
%\end{array}
%\end{aligned}
%\end{equation}
%
%where each element in $E$ is i.i.d zero-mean white Gaussian noise and they are independent.
%\begin{equation}
%\begin{aligned}
%&y_{i}(t)\\
%&=D^y_{i1}(z^{-1})y_{1}(t)+\ldots +[1-N_i(z^{-1})]y_{i}(t)\\
%&+D^u_{im}(z^{-1})u_{m}(t)+e_{i}(t)
%\label{armax}
%\end{aligned}
%\end{equation}
%where $y(t)$ denotes the node, $u(t)$ input and $e(t)$ i.i.d Gaussian noise and:
%\begin{equation}
%\begin{aligned}
%N_i(z^{-1})&=\alpha_{i1}z^{-k}+\alpha_{i2}z^{-k+1}+\ldots +\alpha_{ik}z^{-1}+1\\
%D^y_{ij}(z^{-1})&=\lambda^y_{ij1}z^{-k}+\ldots +\lambda^y_{ij(k-1)}z^{-2}+\lambda^y_{ijk}z^{-1}\\
%D^u_{ij}(z^{-1})&=\lambda^u_{ij1}z^{-k}+\ldots +\lambda^y_{ij(k-1)}z^{-2}+\lambda^u_{ijk}z^{-1}\\
%\end{aligned}
%\end{equation}

Assume the availability of time-series data collected from discrete time indices $1$ to $t$ for each node and input. For the $i$th node, define the following matrices and vectors:
\begin{equation}
\begin{aligned}
&y = \left[\begin{array}{c}y_i(t)\\\vdots \\y_i(k+1)\end{array}\right], w = \left[\begin{array}{c}w_1\\ \hline \vdots \\ \hline w_{p+m}\end{array}\right],\\
&\Phi =\left[\begin{array}{c|c|c} \Phi_1&...&\Phi_{p+m}\end{array}\right],\\
& \lambda = E\{e_i(t)^2\},~ E\{e_i(t)\}=0,\\
&w_r =\left\{\begin{array}{lll} \left[\begin{array}{ccc}a^{ir}_{1} &\ldots& a^{ir}_{k}\end{array}\right]^T, \quad  r\leq p\\
\left[\begin{array}{ccc}b^{i(r-p)}_{1} &\ldots& b^{i(r-p)}_{k}\end{array}\right]^T, p< r\leq p+m \\
\end{array} \right., \\
%& \varepsilon_i(t) = D^e_{ii}(q^{-1})e_{i}(t)\\
%\left[\begin{array}{c|c|c} y_1(t-k:t-1)&...&-y_i(t-k:t-1)\\ \vdots &...& \vdots \\ y_1(1:k)&...&-y_i(1:k)\end{array}\right]\\
&\Phi_{r\leq p}=\left[\begin{array}{c}-y_r(t-k:t-1)\\\vdots\\-y_r(1:k)\end{array}\right],\\
&\Phi_{p<r\leq p+m}=\left[\begin{array}{c}u_{r-p}(t-k:t-1)\\\vdots\\u_{r-p}(1:k)\end{array}\right],
\end{aligned}
\label{para}
\end{equation}
where $y\in R^{t-k}$, $w\in R^{k(p+m)}$ and $\Phi \in R^{(t-k)\times k(p+m)}$.

The vector $w$ is divided into $p$ groups, since there are $p$ nodes in the network. Each group of $w$ contains coefficients of polynomials with respect to a specific node or input. For example, $w_3$ consists of the coefficients of $[A(z^{-1})]_{i3}$ of node $3$. Elements within each group are also indexed so that $w_{rj}$ denotes the $j$th coefficient, $a^{ir}_j$ of $[A(z^{-1})]_{ir}$. If node or input $r$ does not control node $i$, group $w_r$ or $w_{r+p}$ equals $0$. In addition, within each group, the coefficients of excessive polynomial terms are $0$ ($w_{rj}=0$ if $r\leq p$, $j>n_a^*$ or $r> p$, $j>n_b^*$). Consequently, $w$ is both group and element sparse. Since the topology and polynomial orders, $n_a^*$ and $n_b^*$ are unknown, the position of $0$s remains to be determined.

The likelihood distribution based on Bayes' rules is:
\begin{equation}
\begin{aligned}
p(y\big|w,\lambda)&=\frac{1}{(2\pi\lambda)^{(t-k)/2}}\exp\{-\frac{1}{2\lambda}\|y-\Phi w\|_2^2\}.
\end{aligned}
\label{like}
\end{equation}
%To simplify the notation, $Y_k^1$ is ignored in the rest of this paper.

%\begin{equation}
%\begin{aligned}
%p(y_i(t-l)\big|Y_{t-l-1}^{t-l-k},w,\lambda)=\mathcal{N}(y_i(t-l)\big|A(l+1,:)w,\lambda)
%\end{aligned}
%\end{equation}

\begin{rem}For homogeneous dataset collected from multiple experiments subject to the same experimental conditions, the likelihood distribution is the product of that of individual experiment. For heterogeneous dataset, the problem can be formulated in a similar way where some groups of $w$ share the same sparsity profile~\cite{hetero}. The framework of our work can still be applied this case.
\end{rem}

Note that the likelihood distribution is not Gaussian. However, its logarithm is a quadratic function of $w$. Maximizing the likelihood with respect to $w$ leads to PEM or ML methods. With infinite data points, PEM guarantees convergence to the ground truth model~\cite{sys}. In practice, however, with limited data PEM may suffer from over-fitting, typically leading to high order models (nonzero coefficients of excessive polynomials), resulting in fully connected  networks, even if the true one is sparse. An alternative is to penalize for both network topology and model complexity. Referring to parametrizations in~\eqref{para}, a sparse network can be interpreted as group sparse $w$, whereas sparsity within each group indicates reduced order of polynomials. A direct framework to achieve these two levels of sparsity is Sparse Group Lasso. However, Sparse Group Lasso requires parameter tuning. To avoid this limitation, we resort to Sparse Bayesian Learning.

\begin{rem} In practice, most systems are better modelled with ARMAX models instead of just ARX. However, it is not possible to efficiently solve both group and element sparsity with ARMAX models. Nevertheless, as we show below, for ARX models our framework is efficient and has a high degree of freedom in estimating model parameters. In addition, we can easily extend the results to identify NARX models (discussed in Section \ref{ndsf}).
\end{rem}
% eq.~\eqref{armax} can be written as a linear regression model:
% \begin{equation}Y_i=A_iw_i+\Xi_i  \end{equation}
%
%To reconstruct the whole network, we need to solve $p$ independent linear regression problems in the same form due to $p$ nodes. For simplicity, we use a unified expression for all the nodes:
% \begin{equation}y=Aw+\Xi \label{linear} \end{equation}
%
%Now, we aim to solve this linear regression problem given measured data recorded in $Y$ and $A$.
%If the real network topology is sparse and the system order $k$ is set to be much bigger than the true one (since we don't know the true system order, instead, we set an upper bound for it), the weighting vector $w$ is both group-sparse (network is sparse) and element-sparse in each group.

%%%%%%%%%%%%%%%%%%%%%%%%%%%%%%%%%%%%%%%%%%%%%%%%%%%%%%%%%%%%%%%%%%%%%%%%%%%%%%%%%

\section{INDUCING SPARSITY VIA SPARSE BAYESIAN LEARNING}\label{sec:Bayesian}
\subsection{Sparsity inducing priors}
%From a complete Bayesian perspective, all the unknown variables in the linear regression model~\eqref{linear} are treated as stochastic random variables and assigned with a  probability density function (PDF) individually. According to eq.~\eqref{def}, each block in the noise vector $\Xi$ is a random variable following a Multivariate Gaussian distribution. Since these blocks present noise from different experiments, they are independent Gaussian. For eq.~\eqref{armax}, as the noise in each experiment is generated by a Moving-Average (MA) process, the covariance matrix for each block possesses a symmetric Toeplitz structure. As a result, the vector $\Xi\sim \mathcal{N}(0,\Pi)$ itself is Multivariate Gaussian whose covariance matrix has a block diagonal structure with each block to be a symmetric Toeplitz matrix. In this case, the likelihood of $Y$ given the weighting vector $w$ is $p(Y|w)=\mathcal{N}(Y|Aw,\Pi)$.
Full Bayesian methods require a prior distribution for $w$. We define a distribution $p(w)$ in a general form as: $p(w)\propto \exp\left[-\frac{1}{2}\sum_{j}g(w_j)\right]$. Since $w$ is a sparse vector, we assign to $p(w)$ a prior inducing sparsity like Generalized Gaussian, Student's t or Logistic. Such prior functions are usually concave and non-decreasing with respect to $|w_j|$ \cite{prior1}. In this case, however, the estimation of $w$ as the posterior mean is intractable because the posterior distribution $p(w|y)$ is non-Gaussian and not analytical. 

Sparse Bayesian Learning approximates $p(w|y)$ with a Gaussian distribution, so that the estimation $E(w|y)$ can be easily calculated. A sparse inducing prior $p(w)$ is first presented in a variational form, which yields a lower bound, $\hat{p}(w)$, for that prior ($\hat{p}(w)\leq p(w)$) \cite{prior2, graph}. The property of the lower bound is controlled by its hyperparameters. A designed criterion is then applied to  find the best hyperparameters. 

As discussed, the parameter vector $w$ must be both element and group sparse. There are priors able to induce either of these two types of sparsity. We use these priors to construct a novel one that can impose both sparsities simultaneously.

Sparse priors able to induce element sparsity to $w\in R^{k(p+m)}$ can be expressed in the convex type variational from as \cite{statewei, prior1}:
\begin{equation}
\begin{aligned}
&p(w)=\prod^{p+m}_{r=1}p(w_r)=\max_{\beta \geq0}  \mathcal{N}(w | 0,B)\varphi^\beta(\beta),\\
&p(w_r)=\prod_{j=1}^k p(w_{rj})=\max_{\beta _{r}\geq0}  \mathcal{N}(w_r | 0,B_r)\varphi_r^\beta(\beta_{r}), \\
&p(w_{rj})=\max_{\beta _{rj}\geq0}  \mathcal{N}(w_{rj} | 0,\beta_{rj})\varphi_{rj}^\beta(\beta_{rj}).
\label{element}
\end{aligned}
\end{equation}
where subscript $r$ denotes the $r$th group in a vector and $j$ the $j$th element in that group. $\beta=\text{vec}\{\beta_1,...,\beta_{p+m}\}\in R^{k(p+m)}$ is a vector of hyperparameters and $\beta_r=\text{vec}\{\beta_{r1},...,\beta_{rk}\}$. $B$ is the covariance matrix of Gaussian distribution and parametrised by vector $\beta$ as $B=\text{blkdiag}\{B_1,...,B_{p+m}\}$ and $B_r=\text{diag}\{\beta_r\}$. $\mathcal{N}(w|\mu,\Sigma)$ denotes the Gaussian distribution of $w$ with mean $\mu$ and covariance $\Sigma$. $\varphi^\beta_{rj}(\beta)$ is a positive function that depends on the prior $p(w)$. $\varphi^\beta_{r}(\beta)=\prod_{j=1}^k \varphi^\beta_{rj}(\beta_{rj})$ and $\varphi^\beta(\beta)=\prod_{r=1}^{p+m} \varphi^\beta_{r}(\beta_r)$.

%\begin{equation}
%\begin{aligned}
%&w = vec\{w_1\ldots w_{p+m}\},
%%&w = \left[\begin{array}{c}w_1\\ \hline \vdots \\ \hline w_{p+m}\end{array}\right],
%&w_i = vec\{w_{i1}\ldots w_{ik}\}\\
%%w_i = \left[\begin{array}{c}w_{i1}\\  \vdots \\  w_{ik}\end{array}\right]\\
%&\beta = vec\{\beta_1\ldots \beta_{p+m}\},
%%&\beta = \left[\begin{array}{c}\beta_1\\ \hline \vdots \\ \hline \beta_{p+m}\end{array}\right],
%&\beta_i = vec\{\beta_{i1}\ldots \beta_{ik}\}\\
%%\beta_i = \left[\begin{array}{c}\beta_{i1}\\ \vdots \\  \beta_{ik}\end{array}\right]\\
%&B = blkdiag\{B_{1}\ldots B_{p+m}\},
%%&B = \left[\begin{array}{ccc} B_{1}&\ldots &0 \\ \vdots& \ddots & \vdots \\ 0&\ldots &B_{p+m} \end{array} \right]
%&B_i = diag\{\beta_{i1}\ldots \beta_{ik}\}
%%B_i=\left[\begin{array}{ccc} \beta_{i1}&\ldots &0 \\ \vdots& \ddots & \vdots \\ 0&\ldots &\beta_{ik} \end{array} \right].
%\end{aligned}
%\end{equation}
To impose group sparsity, the hyperparameters of each group are unified so that elements in a group share the same sparse profile \cite{heter, intrablock}:
\begin{equation}
\begin{aligned}
&p(w)=\prod^{p+m}_{r=1}p(w_r)=\max_{\gamma\geq0}  \mathcal{N}(w | 0,\Gamma)\varphi^\gamma(\gamma),\\
&p(w_r)=\max_{\gamma_r\geq0}  \mathcal{N}(w_r | 0,\gamma_rI)\varphi_r^\gamma(\gamma_r),
\label{group}
\end{aligned}
\end{equation}
where $\gamma=vec\{\gamma_1,...,\gamma_{p+m}\}$ is a vector of hyperparameters and $\Gamma=\text{blkdiag}\{\gamma_1I,..,\gamma_{p+m}I\}$. $\varphi^\gamma_r(\gamma_r)$ is a positive function and $\varphi^\gamma(\gamma)=\prod_{r=1}^{p+m}\varphi^\gamma_r(\gamma_r)$.
%\begin{equation}
%\begin{aligned}
%&\gamma= vec\{\gamma_{1}\ldots \gamma_{p+m}\}\in\mathbb{R}^{p+m}\\
%%&\gamma=\left[\begin{array}{c} \gamma_{1} \\ \vdots \\\gamma_{p+m} \end{array} \right]\\
%&\Gamma_i= \gamma_{i}I\in\mathbb{R}^{k\times k},
%%\Gamma_i=\left[\begin{array}{ccc} \gamma_{i}&\ldots &0 \\ \vdots& \ddots & \vdots \\ 0&\ldots & \gamma_{i} \end{array} \right]\in\mathbb{R}^{k\times k},\\
%&\Gamma = blkdiag\{\Gamma_{1}\ldots  \Gamma_{p+m}\}
%%&\Gamma=\left[\begin{array}{ccc} \Gamma_{1}&\ldots &0 \\ \vdots& \ddots & \vdots \\ 0&\ldots & \Gamma_{p+m} \end{array} \right].
%\end{aligned}
%\end{equation}
%\begin{rem}
%In practice, if the model is linear ARX, the group of $w$ that presents auto-regression can be excluded from the group prior to improve the estimation accuracy. \jmg{not sure I understood this remark...}
%\end{rem}

According to the construction of element and group sparse priors, neither of them is suitable to impose both kinds of sparsity. The hyperparameters for each group of $w$ in~\eqref{element} are independent so that the resulting lower bound, $\hat{p}(w;\beta)$ is too loose to impose group sparsity. In contrast, the lower bound, $\hat{p}(w;\gamma)$ of~\eqref{group} prohibits the element sparsity within each group thus too rigid. To promote both element and group sparsity, we combine~\eqref{element} and~\eqref{group} to get a new distribution:
\begin{equation}
\begin{aligned}
p(w)=C\max_{\gamma\geq0, \beta\geq0}  \mathcal{N}(w | \varepsilon,B)\mathcal{N}(w | 0,\Gamma)\varphi^{\beta}(\beta)\varphi^{\gamma}(\gamma),
\end{aligned}
\label{GEP}
\end{equation}
where $C$ is the normalization constant that can be absorbed by positive functions $\varphi^{\beta}(\beta)$ or $\varphi^{\gamma}(\gamma)$ and is independent on hyperparameters, $\gamma$ and $\beta$. $\varepsilon$ is the expected value of $w$. As $w$ is element sparse within each nonzero group, $\varepsilon$ is set close to $0$ (e.g. $\parallel \varepsilon \parallel=10^{-3}$). Hence, we get an improper prior as the lower bound of~\eqref{GEP} given by
\begin{equation}
\begin{aligned}
\hat{p}(w)& = \mathcal{N}(w | \varepsilon,B) \mathcal{N}(w | 0,\Gamma)\varphi^{\beta}(\beta)\varphi^{\gamma}(\gamma)\\
&\leq p(w).
\end{aligned}
\label{cprior}
\end{equation}
%If $\gamma_{r}$ approaches 0, the $r$th group of $w$ is enforced to 0 leading to group sparsity. If $\gamma_{r}\neq 0$ and $\beta_{rj}=0$, $w_{rj}=\varepsilon_{rj}$ with probability $1$. In addition, $\gamma$ and $\beta$ cannot be both $0$ or the lower bound is not well defined. 
%In addition, $\varepsilon$ is always set to be non-zero because the multiplication of two impulse functions caused by $\gamma=\beta=0$ is not well defined mathematically. By setting $\lvert\varepsilon\rvert$ small enough, the prior is able to promote element sparsity.
The prior in~\eqref{cprior} shows that two types of sparsity are controlled by two series of hyperparameters, $\beta$ and $\gamma$ respectively. As $\gamma_r$ approaches $0$, the $r$th group of $w$ is enforced to $0$ regardless of $\beta_r$. That means the group sparsity can be determined from the hyperparameter space of dimension $p+m$ instead of $k(p+m)$ when only element sparse inducing priors are applied. Furthermore, within a nonzero group ($\gamma_r\neq 0$), hyperparameter $\beta_{r}\in R^k$ enables extra freedom to search for each element of $w_{r}$ whilst GSBL only allows one degree of freedom to tune $w_r$ via $\gamma_r\in R$. The values of these hyperparameters are unknown and remain to be estimated from the data.

%As a result, for a target vector consisting of $q$ non-zero groups out of $p$ in total and $k$ elements in each group, its sparse profile is determined from a $\{R^p\}\times \{R^{qk}\}$ hyperparameter space compared to  $\{R^{(p-q)k}\}\times \{R^{qk}\}$ without group sparse priors.

\begin{rem}The conventional way to promote both element and group sparsity is to use hierarchical Bayesian by introducing two hyperparameters where one is conditioned on the other. However, the hyperparameter that is deeper in the hierarchy has less impact on the inference procedure \cite{t2}. This means that the resultant penalty cannot impose element and group sparsity at the same time. Instead, multiplying two priors results in both hyperparameters influencing $w$  directly. 
\end{rem}

\subsection{Type II maximization}
Although the prior $\hat{p}(w)$ is improper, we can still get a normalized posterior distribution of $w$ as:
\begin{equation}
\begin{aligned}
\hat{p}(w|y)&= \frac{p(y|w)\hat{p}(w)}{\int p(y|w)\hat{p}(w)dw}.
\end{aligned}
\end{equation}

Clearly, $\hat{p}(w|y)$ is a Gaussian distribution as its logarithm is quadratic with respect to $w$:
\begin{equation}
\begin{aligned}
\hat{p}(w|y)= \mathcal{N}(w|\mu,\Sigma),
\end{aligned}
\label{pw1}
\end{equation}
where
\begin{equation}
\begin{aligned}
\Sigma&=\left[(\Gamma^{-1}+B^{-1})+\lambda^{-1}\Phi^T\Phi\right]^{-1},\\
\mu&=\Sigma(\lambda^{-1}\Phi^Ty-B^{-1}\varepsilon).
\end{aligned}
\label{pw2}
\end{equation}
Since the true posterior distribution, $p(w|y)$ is intractable, SBL employs $\hat{p}(w|y)$ as the approximation. The sparsity of the estimated $w$ as $E(w|y)=\mu$ depends on hyperparameters $\beta$ and $\gamma$. SBL introduces empirical Bayes to estimate their optimal values.

Hyperparameters $\beta$ and $\gamma$ are selected so that $\hat{p}(w|y)$ is close to the true posterior distribution, $p(w|y)$ under the designed criterion. One way is to minimize the misaligned mass between $p(w)$ and $\hat{p}(w)$ weighted by the marginal likelihood $p(y|w)$, which is called evidence maximisation or Type II maximisation \cite{prior1, pattern, statewei}. It is equivalent to estimating hyperparameters using the maximum likelihood method:
\begin{equation}
\begin{aligned}
(\gamma^* ,\beta^*, \lambda^*)&=\arg \min_{\beta, \gamma,\lambda\geq0}\int p(y|w)|p(w)-\hat{p}(w)|dw \\
%&= \arg \max_{\beta, \gamma,\lambda\geq0} \int p(y|w)\hat{p}(w)dw \\
&=\arg \min_{\beta, \gamma,\lambda\geq0} -2\log\int p(y|w)\hat{p}(w)dw \\
&=\arg \min_{\beta, \gamma,\lambda\geq0} -2\log~\hat{p}(y|\beta,\gamma,\lambda).
\end{aligned}
\label{opt}
\end{equation}
\begin{rem}Not all sparse inducing priors can lead to a sparse solutions under the framework of SBL. The selection of functions $\varphi^\beta(\cdot)$ and $\varphi^\gamma(\cdot)$ influences the sparsity of the final result. It has been shown that one reasonable option is to make $-\log\varphi(\cdot)$ concave and nondecreasing~\cite{prior1}. Instead, this paper sets $\varphi(\cdot)$ as a constant, which means the original prior $p(w)$ is a Student's t distribution. As a result, the function  $\varphi(\cdot)$ can be ignored in the following discussion.
\end{rem}
%\noindent
%\begin{rem}: $\Pi$ matrix has a block-diagonal symmetric Toeplitz structure. It is parameterized by $\frac{mk(2M-k+1)}{2}$ variables. Estimation of $\Pi$ is very important because the recovery performance can be poor if estimated $\Pi$ is sub-optimal \cite{intrablock}. To avoid over-fitting and reduce computation complexity, we assume 1) white Gaussian noise $e(t)$ in each experiment has same variance (number of variables reduces to $\frac{k(2M-k+1)}{2}$). 2) Each diagonal block of $\Pi$ is well approximated by its main diagonal elements. Consequently, $\Pi$ is only parameterized by one variable so that $\Pi=\sigma^2I$. Simulation indicates such approximation is reasonable and still leads to good performance.
%\end{rem}

\begin{prop} The estimation of $w$ as the posterior mean in~\eqref{pw2} can be incorporated into the Type II maximization~\eqref{opt}, thus resulting in the optimisation problem as follows:
	\begin{equation}
	\begin{aligned}
	&\mathcal{L}: \min_{\beta , \gamma,\lambda,w}\lambda^{-1}\|y-\Phi w\|_2^2+\|w\|^{2}_{\Gamma^{-1}} +\|w-\varepsilon\|^{2}_{B^{-1}}\\
	&+\log|B+\Gamma|+\log|\lambda I+\Phi(\Gamma^{-1}+B^{-1})^{-1}\Phi^T|\\
	&\textrm{subject to:}\\
	& \quad \quad \quad \quad \quad \quad  \beta\geq0 , \gamma\geq0, \lambda\geq0.
	%&\textrm{Or:}\\
	%&\mathcal{L}_2: \\
	%&\min_{\beta , \gamma,\lambda}y^T\left[\lambda I+\Phi(\Gamma^{-1}+B^{-1})^{-1}\Phi^T\right]^{-1}y\\
	%&+\log|B+\Gamma|+log|\lambda I+\Phi(\Gamma^{-1}+B^{-1})^{-1}\Phi^T|\\
	%&\textrm{Subject to:}\\
	%& \quad \quad \quad \quad \quad \quad  \beta\geq0 , \gamma\geq0, \lambda\geq0
	\end{aligned}
	\label{opt1}
	\end{equation}
\end{prop}

\begin{IEEEproof} The derivation can be found in Appendix~\ref{sec:app1}.
\end{IEEEproof}

\begin{rem}The advantage of embedding the estimation of $w$ into~\eqref{opt} is that the resulting optimisation problem can be decomposed into sub-problems. This way, a large-scale problem can be solved more efficiently. The decomposition is discussed in the following sections. In addition, the resulting algorithm interprets the mechanism of the proposed method to impose both kinds of sparsity.
\end{rem}
%\begin{rem}:
%The optimization problem~\eqref{opt1} can also be derived using hierarchical Bayesian format. Assume $p(w|L)= \mathcal{N}(w|0,L)$, $p(L|P)=(\frac{2}{3})^{p+m}|P|^{\frac{3}{2}}|P-L|^{1/2}$ and $p(P)\propto|P|^{-\frac{5}{2}}$ where $L$ and $P$ are counterparts of $B$ and $\Gamma$ respectively, then $p(w)\propto\int p(w|L)p(L|P)p(P)dLdP$. In SBL, normalized $p(w|y,L,P)$ is used to approximate $p(w|y)$. The value of $L$ and $P$ is evaluated by solving MAP of marginal distribution of hyperparameters as $\max p(L,P|y)$. After simple manipulation, the resultant optimization problem is same with~\eqref{opt1}. In this hierarchical format, $L$ takes charge of element sparsity of $w$. $p(L|P)$ brings an underlying constraint $0\preceq L\preceq\Gamma$ which controls sparsity of $L$. $p(P)$ is similar to a Jeffrey prior thus enforcing group sparsity to $L$. Therefore, hyperparameter $P$ imposes group sparsity to $w$ indirectly via $L$.
%\end{rem}

\subsection{Algorithm to solve Type II maximization}
Although the optimisation problem $\mathcal{L}$ is nonlinear and nonconvex, we can formulate the cost function as a difference of two convex terms and then solve the problem as a Difference of Convex Programming (DCP). In addition, we prefer to employ distributed algorithms because, in practice, the inference problem often has to deal with large scale networks and enormous datasets. It turns out that DCP can be further decomposed using the Alternating Direction Method of Multipliers (ADMM), which makes it possible to tackle large scale networks and utilizes limited computational power more efficiently.

\subsubsection{Solve $\mathcal{L}$ using Convex-Concave Procedure}
The cost function is separated into two parts in the following way:

\begin{equation}
\begin{aligned}
\mathcal{L}: \min_{\gamma\geq0, \beta\geq 0, \lambda\geq0}
u(w,\lambda,\beta,\gamma)-v(\lambda,\beta,\gamma),
\end{aligned}
\label{update}
\end{equation}
where
\begin{equation}
\begin{aligned}
&u=\lambda^{-1}\|y-\Phi w\|_2^2+\|w\|^{2}_{\Gamma^{-1}} +\|w-\varepsilon\|^{2}_{B^{-1}},\\
&v=-\log|B+\Gamma|-\log|\lambda I+\Phi(\Gamma^{-1}+B^{-1})^{-1}\Phi^T|.
\end{aligned}
\end{equation}

\begin{prop}: Functions $u(w,\lambda,\beta,\gamma)$ and $v(\lambda,\beta,\gamma)$ are both jointly convex with respect to their own variables.\end{prop}

\begin{IEEEproof}The derivation can be found in Appendix~\ref{sec:app2}.
\end{IEEEproof}
Now, the optimisation problem is transferred into a difference of convex programming (DCP). It can be solved using sequential convex optimisation techniques. Here, we use a convex-concave procedure (CCCP), which is a type of majorisation-minimisation (MM) algorithm using the linear majorisation function \cite{nonp3, cccp}. For $\min_xf(x)$  where $f(x)=u(x)-v(x)$, and $u(x)$ and $v(x)$ are convex, we can solve it iteratively by:
\begin{equation}
\begin{aligned}
x^{n+1}=\arg\min_{x}u(x)-<x,\nabla v(x^{n})>,
\end{aligned}
\end{equation}
where $<\cdot,\cdot>$ denotes inner product.

Therefore, in each iteration, we solve a convex problem:
\begin{equation}
\begin{aligned}
&\left[\gamma^{n+1},\beta^{n+1}, \lambda^{n+1}, w^{n+1}\right]\\
&=\arg\min_{\gamma, \beta, \lambda,w}
u(w,\lambda,\beta,\gamma)-\frac{\partial v}{\partial\lambda}\big|_{(\lambda^{n},\beta^{n},\gamma^{n})}\lambda\\
&-\nabla^T_{\beta}v\big|_{(\lambda^{n},\beta^{n},\gamma^{n})}\beta-\nabla^T_{\gamma}v\big|_{(\lambda^{n},\beta^{n},\gamma^{n})}\gamma\\
&\textrm{subject to:}\\
& \quad \quad \quad \quad \quad \quad  \beta\geq0 , \gamma\geq0, \lambda\geq0,
\end{aligned}
\label{1cccp}
\end{equation}
where
\begin{equation}
\begin{aligned}
-\frac{\partial v}{\partial\lambda}&=trace\{\Delta\}\\
-[\nabla_{\beta}v]_{rj}&=\frac{1}{(\gamma_r+\beta_{rj})}+ \frac{\gamma_r^2\left[\Phi^T\Delta \Phi \right]_{qq}}{(\gamma_r+\beta_{rj})^{2}} \\
-\frac{\partial v}{\partial\gamma_r}&=\sum_{j=1}^{k}\frac{1}{\gamma_r+\beta_{rj}}+\frac{\beta_{rj}^2\left[\Phi^T\Delta \Phi\right]_{qq}}{(\beta_{rj}+\gamma_r)^2}\\
\Delta &= [\lambda I+\Phi(\Gamma^{-1}+B^{-1})^{-1}\Phi^T]^{-1}\\
q&=(r-1)k+j.
\end{aligned}
\label{1up}
\end{equation}
If we optimize $\gamma$, $\beta$ and $\lambda$ first, we get analytical expressions of their optimal solutions as functions of $w$:
\begin{equation}
\begin{aligned}
\beta_{rj}^{opt}&=\frac{|w_{rj}-\varepsilon_{rj}|}{\sqrt{g_{rj}^{\beta}}},~
\gamma_{r}^{opt}&=\frac{\|w_{r}\|_2}{\sqrt{g_{r}^{\gamma}}},~
\lambda^{opt}&=\frac{\|y-\Phi w\|_2}{\sqrt{g^{\lambda}}},
\end{aligned}
\label{1sol}
\end{equation}
where
\begin{equation}
\begin{aligned}
g_{rj}^{\beta}&=-\left[\nabla_{\beta}v|_{(\lambda^{n},\beta^{n},\gamma^{n})}\right]_{rj}\\
g_{r}^{\gamma}&=-\left[\nabla_{\gamma}v|_{(\lambda^{n},\beta^{n},\gamma^{n})}\right]_r\\
g^{\lambda}&=-\frac{\partial v}{\partial\lambda}|_{(\lambda^{n},\beta^{n},\gamma^{n})}.
\end{aligned}
\label{2up}
\end{equation}
It is easy to see that such solutions are valid since they satisfy the constraint. Therefore, sub-problem (\ref{1cccp}) can be further simplified by substituting (\ref{1sol}):
\begin{equation}
\begin{aligned}
w^{n+1}&=\arg\min_{w}\sqrt{g^{\lambda}}\|y-\Phi w\|_2\\
& +\sum_{r=1}^{p+m}\left(\sqrt{g_{r}^{\gamma}}\|w_r\|_2 +\sum_{j=1}^{k} \sqrt{g_{rj}^{\beta}}|w_{rj}-\varepsilon_{rj}|\right).
\end{aligned}
\label{glasso}
\end{equation}

The optimisation~\eqref{glasso} can be solved as a Second Order Cone Program (SOCP). \eqref{glasso} is a reweighted LASSO type problem with a minor variant in that the first term of data-fitting error is measured by $\ell_2$-norm rather than by sum of squares. Without estimating the noise variance, $\lambda$,~\eqref{glasso} can be reformulated into a standard SGL algorithm. In addition, the second term of the cost function blends $\ell_1$ and $\ell_2$ penalties of $w$ to impose element and group sparsity at the same time. The weights of these two terms are updated automatically in each iteration, thus avoiding extra tuning.
%  It has a very similar form with SGL which blends $\ell_1$ and $\ell_2$-norm as the penalty~\cite{SGL}. The difference is an extra weight imposed to the first data-dependent term with $\lVert \cdot \rVert_{2}^2$ instead of $\lVert \cdot \rVert_{2}$ thus leading to a very different function property. Such variation is due to the estimation of $\lambda$ incorporated into the problem. If $\lambda$ is known, the resulting optimization problem can be reformulated as a reweighted SGL.

Special attention should be paid to the estimation of the noise variance, $\lambda$~\cite{tipp,emSBL}. Under some circumstances, $\lambda$ can be compensated by hyperparameters. The risk decreases as more data points are available for identification. The estimated noise variance has a significant impact on the final result~\cite{intrablock}. If the algorithm yields an abnormal value, it is reasonable to resort to other criteria. One conventional way is cross-validation; the other is to fit the data with an ARX model of high order and use the prediction error as the approximation of the sampled noise~\cite{nonp}. The empirical noise variance can then be calculated from the error.
%  For sake of demonstrationimatio, assume each group of $w$ only has $1$ element and $A=[\hat{A},I]$. To achieve the optimal value of cost function in $\mathcal{L}_2$ given by $\lambda_{opt}$, $\beta_{opt}$ and $\gamma_{opt}$, we have to make sure that there exists $\lambda\neq\lambda_{opt}$ such that the following problem has feasible solutions:
%\begin{equation}
%\begin{aligned}
%&(x^{-1}+y^{-1})^{-1}=k_1-\lambda\\
%&x+y=k_2,
%\end{aligned}
%\label{id}
%\end{equation}
%where $k_1$ and $k_2$ are fixed by optima. It is easy to see that as long as $k_2^2-(k_1-\lambda)k_2\geq0$, there is at least one feasible solution so that $\lambda$ is unidentifiable.

To summarize, Algorithm~\ref{alg1} represents the procedure above.

%\begin{alg}\label{alg1}
% CCCP to solve $\mathcal{L}_1$\\
%1: Initialize $\beta^0$, $\gamma^0$, $(\lambda)^0$\\
%2: Calculate $\alpha^3$, $\alpha_{i}^2$ and $\alpha_{ij}^1$ using (\ref{1up}) and (\ref{2up})\\
%3: For $n=1:Max$ do\\
%4: Solve the reweighted Variational Sparse Group Lasso problem:\\
%$\left[w^{n+1}\right]=\arg\min_{w}$\\
%$\sqrt{\alpha^3}\|y-Aw\|_2+ \sum_{i=1}^{p+m}\sum_{j=1}^{k}\sqrt{\alpha_i^2}\|w_i\|_2 + \sqrt{\alpha_{ij}^1}|w_{ij}|$\\
%5: Update $\beta^{n+1}$, $\gamma^{n+1}$ and $(\lambda)^{n+1}$ according to (\ref{1sol})\\
%6: Prune out small $\beta_{ij}$ and $\gamma_{i}$ and remove corresponding columns of dictionary matrix $A$\\
%7: Update $\alpha^3$, $\alpha_{i}^2$ and $\alpha_{ij}^1$ using (\ref{1up}) and (\ref{2up})\\
%8: If any stopping criteria is satisfied, break\\
%9: end for
%\end{alg}

\begin{algorithm}[!]
	\caption{Solve $\mathcal{L}$ using CCCP}
	\label{alg1}
	\begin{algorithmic}[1]
		\State Initialize $\beta^0$, $\gamma^0$, $\lambda^0$
		\State Calculate $g_{rj}^{\beta}$, $g_{r}^{\gamma}$ and $g^{\lambda}$ using (\ref{1up}) and (\ref{2up})
		\For  {$n=1:Max$}
		\State Solve the reweighted convex problem~\eqref{glasso}
		\State Update $\beta^{n+1}$, $\gamma^{n+1}$ and $\lambda^{n+1}$ according to (\ref{1sol})
		\State Update $g_{rj}^{\beta}$, $g_{r}^{\gamma}$ and $g^{\lambda}$ using (\ref{1up}) and (\ref{2up})
		\If {Any stopping criteria is satisfied}
		\State Break
		\EndIf
		\EndFor
	\end{algorithmic}
\end{algorithm}

\subsubsection{Decompose the optimization by Alternating Direction Method of Multipliers (ADMM)}

If a network possesses a large number of nodes, leading to high dimensional variables, solving (\ref{glasso}) directly will be very inefficient, or even computationally infeasible, due to hardware limitations. In this case, we split the optimisation problem into a series of small-scale sub-problems to reduce computational burden so that they can be coped with in parallel. It turns out that by splitting the cost function, \eqref{glasso} can be formulated as a sharing problem and solved using ADMM algorithms \cite{admm, admmwei}.

A sharing problem is presented in the form: $\min \sum f_r(x_r) + g(\sum x_r)$ where $x_r$ denotes the $r$th group of vector $x$, and $g(\cdot)$ and $f_r(\cdot)$ are convex. Problem \eqref{glasso} can be rewritten as a sharing problem in ADMM form: 
\begin{equation}
\begin{aligned}
w^{n+1}&=\arg\min_{w}\sqrt{g^{\lambda}}\|y-\sum^{p+m}_{r=1}z_r\|_2\\
&+\sum_{r=1}^{p+m}\left(\sqrt{g_{r}^{\gamma}}\|w_r\|_2 +\sum_{j=1}^{k} \sqrt{g_{rj}^{\beta}}|w_{rj}-\varepsilon_{rj}|\right),\\
\textrm{subject to:}\\
&\quad\quad\quad\quad\quad \Phi_rw_r - z_r = 0,
\end{aligned}
\label{sharing}
\end{equation}
where $\Phi = \left[\begin{array}{c|c|c}\Phi_1&\ldots&\Phi_{p+m}\end{array}\right]$ as in~\eqref{para}.

The scaled form of~\eqref{sharing} is \cite{admm}:
\begin{equation}
\begin{aligned}
w^{n+1}_r=&\arg\min_{w_r}\sqrt{g_{r}^{\gamma}}\|w_r\|_2 +\sum_{j=1}^{k}\sqrt{g_{rj}^{\beta}}|w_{rj}-\varepsilon_{rj}|,\\
& +(\rho/2)\| \Phi_rw_r-\Phi_rw^n_r + \overline{\Phi w}^n - \overline{z}^n+u^n\|^2_2,\\
\overline{z}^{n+1} = &\arg\min_{\overline{z}}\sqrt{g^{\lambda}}\|y-(p+m)\overline{z}\|_2\\
&+\left[(p+m)\rho/2\right]\|\overline{z}-u^n-\overline{\Phi w}^{n+1}\|^2_2,\\
u^{n+1}  =&u^n+\overline{\Phi w}^{n+1}-\overline{z}^{n+1}.
\end{aligned}
\end{equation}
where
\begin{equation}
\begin{aligned}
\overline{\Phi w}^n = \frac{\sum^{p+m}_{r=1}\Phi_rw_r^n}{p+m}.
\end{aligned}
\end{equation}
The original optimisation problem of $w\in R^{k(p+m)}$ is now decomposed into a series of sub-problems, each of which scales as $w_r\in R^{k}$. Hence, the computational burden per iteration is greatly relieved. $w$-update is a standard SGL problem that can be efficiently solved using accelerated generalized gradient descent algorithm~\cite{SGL}. $\overline{z}$-update is a group LASSO problem and has analytical solutions. Let $\hat{z}=\overline{z}-\frac{y}{p+m}$, the original $\overline{z}$-update becomes:
\begin{equation}
\begin{aligned}
\hat{z}^{n+1} = &\arg\min_{\hat{z}}\frac{1}{2}\|\hat{z}+\frac{y}{p+m}-u^n-\overline{\Phi w}^{n+1}\|^2_2\\
&+ \frac{\sqrt{g^{\lambda}}}{\rho}\|\hat{z}\|_2,
\end{aligned}
\end{equation}
so that
\begin{equation}
\begin{aligned}
\hat{z}^{n+1} = \frac{c}{\|c\|_2}\bigg(\|c\|_2-\frac{\sqrt{g^{\lambda}}}{\rho}\bigg)_{+},
\end{aligned}
\end{equation}
where
\begin{equation}
\begin{aligned}
c &= -\frac{y}{p+m}+u^n+\overline{\Phi w}^{n+1}.
\end{aligned}
\end{equation}
$w$-update can be solved in parallel independently. $\overline{z}$ and $u$-update are then solved in sequence after collecting $w$-update.
\subsubsection{Solve $\mathcal{L}$ using Expectation Maximization}
The EM method is a traditional technique to solve~\eqref{opt}. It belongs to the class of majorisation-minimisation (MM) methods and is a special case of DCA (Difference of Convex functions Algorithm). Whilst a DC function has infinite many DC decompositions, the way to decompose the function can greatly influence the performance of the algorithm~\cite{DC}. 

To maximise a likelihood function, $L(\theta)=\log p(y|\theta)$, EM implements Expectation (E step) and Maximisation (M step) iteratively. In the E step, the function, $Q(\theta,\theta^{n})=E_{x|y,\theta^n}[\log p(y,x|\theta)]=\int \log p(y,x|\theta)p(x|y,\theta^n)dx$ is calculated where $x$ is the unobservable latent random variable. In the M step, the optimisation problem, $\theta^{n+1}=\arg\max Q(\theta,\theta^n)$ is solved~\cite{emSBL,pattern}. The generated sequence, $\{\theta^n\}$ leads to the increased likelihood function ($L(\theta^n)<L(\theta^{n+1})$). In our case, we regard $w$ as the latent variable. Following the standard procedure of the EM method, the algorithm is described in Algorithm $2$. 

%To solve the optimisation problem~\eqref{opt} with the matrix $\Phi \in\mathbb{R}^{N\times M}$ and $N\ll M$, the cost of EM method is $O(MN^2)$ in each iteration.

%Another traditional method to solve (17) is to use Expectation Maximization (EM) algorithm, since (17) can be treated as a ML problem with $\hat{p}(w)$ to be the prior. We regard $w$ as the hidden variable~\cite{emSBL,pattern}. Following the standard procedure of EM method, the algorithm is described in Algorithm $2$.

%\begin{alg}\label{alg2}
%Solve $\mathcal{L}_2$ using EM\\
%1: Initialize $\beta^0$, $\gamma^0$, $(\lambda)^0$\\
%2: For $n=1:Max$, do\\
%3: E step: Formulate $p(w|y,\beta^{n},\gamma^{n},(\lambda)^{n})$ according to \eqref{pw1} and \eqref{pw2}\\
%4: M step: \\
%$[\beta^{n+1},\gamma^{n+1},(\lambda)^{n+1}]$\\
%$=\arg\min E_{w|\beta^{n},\gamma^{n},(\lambda)^{n}}[lnp(y,w|\beta,\gamma,(\lambda))]$\\
%5: Update solutions of M step as:\\
%$\gamma_{i}^{n+1}=\frac{1}{k}\sum_{j=1}^{k}\Sigma_{qq}^{n}+(\mu_{ij}^n)^{2}$\\
%$\beta_{ij}^{n+1}=\Sigma_{qq}^{n}+(\mu_{ij}^n)^{2}$\\
%$(\lambda)^{n+1}=\frac{\|y-A\mu^n\|_2^2+(\lambda)^{n}\sum_{i=1}^{p+m}\sum_{j=1}^{k}1-[(\beta_{ij}^n)^{-1}+(\gamma_{i}^n)^{-1}]\Sigma_{qq}^{n}}{N}$\\
%where $q=(i-1)k+j$ and $N=k(p+m)$
%\end{alg}

\begin{algorithm}[!]
	\caption{Solve $\mathcal{L}$ using EM}
	\label{alg2}
	\begin{algorithmic}[1]
		\State Initialize $\beta^0$, $\gamma^0$, $\lambda^0$
		\For  {$n=1:Max$}
		\State E step: Formulate $p(w|y,\beta^{n},\gamma^{n},\lambda^{n})$ according to \eqref{pw1} and \eqref{pw2}
		\State M step: Formulate the optimisation problem and update solutions as:
		\begin{equation}
		\begin{aligned}
		&[\beta^{n+1},\gamma^{n+1},\lambda^{n+1}]\\
		&=\arg\min E_{w|\beta^{n},\gamma^{n},\lambda^{n}}\left\{\ln p(y,w|\beta,\gamma,\lambda)\right\}
		\end{aligned}
		\end{equation}
		\begin{equation}
		\begin{aligned}
		&\gamma_{r}^{n+1}=\frac{1}{k}\sum_{j=1}^{k}[\Sigma^n]_{qq}+(\mu_{rj}^n)^{2}\\
		&\beta_{rj}^{n+1}=\left[\Sigma^n\right]_{qq}+(\mu_{rj}^n-\varepsilon_{rj})^{2}\\
		&\lambda^{n+1}=\frac{\|y-\Phi \mu^n\|_2^2+\lambda^{n}\sum_{r=1}^{p+m}\sum_{j=1}^{k}1-\tau_{rj}[\Sigma^n]_{qq}}{N},
		\end{aligned}
		\end{equation}
		where $\tau_{rj}=(\beta_{rj}^n)^{-1}+(\gamma_{r}^n)^{-1}$, $q=(r-1)k+j$ and $N=k(p+m)$
		\If {Any stopping criteria is satisfied}
		\State Break
		\EndIf
		\EndFor
	\end{algorithmic}
\end{algorithm}
%$\mathcal{L}_2$ can also be solved directly by setting the gradient of cost function $0$ in order to find stationary points~\cite{tipp}:
%\begin{equation}
%\begin{aligned}
%&-\frac{1}{\beta_{ij}^2}(\mu_{ij}^2+\Sigma_{qq})+\frac{1}{\beta_{ij}}=0\\
%&-\frac{1}{\gamma_i^2}\sum_{j=1}^{k}(\mu_{ij}^2 + \Sigma_{qq})+ \frac{k}{\gamma_i}=0\\
%&\lambda=\frac{\|y-\Phi \mu\|_2^2+\lambda\sum_{i=1}^{p+m}\sum_{j=1}^{k}1-(\beta_{ij}^{-1}+\gamma_{i}^{-1})\Sigma_{qq}}{N}
%\end{aligned}
%\label{der}
%\end{equation}
%
%  There is no analytical solution to equations~\eqref{der}. However, we can search for optima by cycling through variables in sequence. Applying simple manipulations to \eqref{der} results in the same iteration in Algorithm~\ref{alg2}.

%\begin{rem}
%$\mathcal{L}_2$ can also be solved using CCCP. The first term of the cost function is convex since it is partial minimum of $u(w,\beta,\gamma)$. Clearly, it coincides with Algorithm~\ref{alg1}. The only difference comes from the sequence of optimizing variables.
%\end{rem}
The cost of EM per iteration is dominated by the inversion of the covariance matrix $\Sigma$ in~\eqref{pw2}. At first glance, the work required seems huge ($O[k^3(p+m)^3]$). Nevertheless, after applying the Sherman\textendash Morrison\textendash Woodbury formula, the cost is reduced to $O([k(p+m)(t-k)^2])$. This merit is lost if the covariance matrix of priors is full (i.e. other kernel functions are used to build the covariance matrix). Consequently, the total cost per iteration is $O[k(p+m)(t-k)^2]$ as, in practice, the scale of the target network is usually large and the measured data are limited ($t-k\ll k(p+m)$). 

\section{EXTENSION TO NONLINEAR ARX MODELS}\label{ndsf}

In reality, most networks are nonlinear (e.g. genetic regulation networks). To cope with these networks, we introduce nonlinear terms into a multivariable ARX model.

Considering a network with $p$ nodes and $m$ inputs: $A(z^{-1};w)Y(t) =  B(z^{-1};w)U(t) + F(t;w) + E(t)$, where the form of matrices $A(z^{-1};w)$ and $B(z^{-1};w)$ is exactly the same with~\eqref{pARX}.  $F(t;w)$ is a vector of nonlinear functions, depending on the past values of nodes and inputs. Each element of $F(t;w)$ is the linear combination of basis functions. The topology of the network is reflected by the nonzero elements in $A(z^{-1};w)$, $B(z^{-1};w)$, and nonlinear terms of $F(t;w)$, whereas the system dynamics is dominated by the elements in these matrices.

We parametrize each node of the network in the same way as~\eqref{armax}: 
\begin{equation}
\begin{aligned}
&y_{i}(t)=-[A(z^{-1})]_{i1}y_{1}(t)-\ldots +\{1-[A(z^{-1})]_{ii}\}y_{i}(t)\\
&+[B(z^{-1})]_{i1}u_{1}(t)...+[B(z^{-1})]_{im}u_{m}(t)+F_i(t)+e_{i}(t)
\label{narx}
\end{aligned}
\end{equation}
where
\begin{equation}
\begin{aligned}
&F_{i}(t)=\sum^p_{r=1}\sum_{j=1}^lc^{ir}_{j}f^{ir}_{j}(t)\\
&f_j^{ir}(t)=g^{ir}_{j}[y_r(1:t-1),u(1:t-1)]
\end{aligned}
\label{nd}
\end{equation}
$F_i(\cdot)$ is the linear combination of nonlinear basis functions $g(\cdot)$, which depends on the past evolution. The coefficient vector $c^i$ is divided into $p$ groups, each of which represents the regulation from a node.  Within each group, there are $l$ elements corresponding to $l$ basis functions.

Vector $c^i$ is group sparse because some nodes do not control the $i$th node (i.e. network is sparse). In addition, it is also element sparse since within each group, only a few nonlinear terms are appropriate to describe the dynamics of the network. For instance, a group of nonlinear terms (e.g. Hill functions) are used to present the potential transcription activity of a transcriptional factor associated to a specific node. The group sparsity determines if this node regulates the target. Besides, only a specific type of the Hill functions in this group is suitable, depending on whether such a transcription is repressive or active.

The estimation of model parameters can be formulated into the same form of~\eqref{para} where $w$ is both element and group sparse. Therefore, the discussed framework in this paper follows.

\section{SIMULATION}\label{sec:Simulation}

To illustrate and test GESBL, simulations were carried out under three cases: linear ARX models, the nonlinear state space model of the three-gene repressilator~\cite{nature}, and the nonlinear state space model of the \textit{Arabidopsis} circadian clock~\cite{Millar10}. Linear ARX models include networks of different topologies: random sparse topology and ring structures. The performance was compared with other methods including the kernel method \cite{nonp}, SBL~\cite{prior1} and GSBL~\cite{heter}. 

Inferred topologies  are evaluated using standard tools: True Positive Rate (TPR),  Precision (Prec), and the percentage of successful inference ($100\%$ TPR and $100\%$ Prec) among all runs. TPR reveals the percentage of how many true links of the ground truth networks are identified. Prec indicates the reliability of inferred networks, which equals ${TP}/{(TP+FP)}$, where TP is the number of true links correctly identified and FP is the number of those incorrectly identified. For example, if Prec is $50\%$, it means that half of the links in the estimated network are wrong.

To evaluate the accuracy of estimated parameters, we calculate the normalised root mean square error (NRMSE) as:
\begin{equation}
\begin{aligned}
NRMSE &= \frac{1}{\sqrt{N}\hat{w}}\|w_{est}- w_{true}\|_{2}\\
\end{aligned}
\end{equation}
where
\begin{equation}
\begin{aligned}
\hat{w} &= \frac{1}{N}\|w_{true}\|_{1}
\end{aligned}
\end{equation}

\subsection{Multivariable ARX models}
Data were simulated from stable and sparse ARX models with $10$ nodes. While all the examples here are stable, GESBL applies to other unstable networks as well, since there are no constraints on system stability.

We conducted two Monte Carlo simulations. The first simulated $100$ networks whose topology and internal dynamics up to $5$th order were generated randomly. Each node was independently driven by an input, so that $B(z^{-1})$ in~\eqref{pARX} is diagonal. The second simulation generated 100 new random networks, this time from a fixed topology: a ring network. There was only one input applied to a single node.

For the first set of simulations, for each random network we generated a $10\times10$ sparse polynomial matrix $A(z^{-1})$. The generated networks included at least one feedback loop. Sparsity of the matrix was controlled by a predefined variable that indicated the probability of $[A(z^{-1})]_{ij}$ to be nonzero. Each polynomial entry of $A(z^{-1})$ was obtained by the function $drmodel$ in Matlab to ensure its roots were constrained to the open unit circle. Polynomial orders were selected from $1$ to $5$ uniformly. Closed loop stability was verified by computing the poles of $A^{-1}(z^{-1})$. The matrix was discarded if it was unstable (any pole is outside the unite circle of the complex plane). The parameters of polynomial matrices $B(z^{-1})$ were then generated using function $randn$ in Matlab and polynomial orders were selected the same way as in $A(z^{-1})$. The above procedure was repeated to generate $100$ models. 

On average, there were about 36 links in each network (out of 90 possible). Both the exciting input (known) and process noise (unknown) were independent Gaussian. Models were simulated with different Signal-Noise Ratio (SNR) and $100$ time-series data points for each node, where SNR is defined as $SNR=10log_{10}({\sigma^2_{u}} / {\sigma^2_{e}})$ and $\sigma^2$ denotes the signal variance. The polynomial order, $k$ was set to $8$ (since the ground truth order was treated unknown during the inference procedure). 

Table~\ref{t1} compares the inferred Boolean network structure using different methods. It indicates that when the process noise is low, all methods are capable of ruling out redundant correlation among nodes ($100\%$ Prec). Our method not only guarantees high TPR but also reconstructs the perfect ground truth topology $77$ times out of $100$. When noise levels rise, the performance of all methods degrades. With relatively small noise ($10dB$), all methods are able to identify most of the links in the network ($TPR>80\%$). However, Prec of SBL is only $59.6\%$, indicating potential high False Positive and low confidence of estimation while the other approaches still retain satisfactory performance. When the noise is high, TPR of SBL and GSBL stay high whereas ours and the kernel method miss approximately half of the ground truth links. Nonetheless, Prec of our method ($99.6\%$) indicates that most of the estimated links are correct. Although SBL and GSBL attain high TPR, due to their low Prec, it is difficult to tell which links are true without prior knowledge of the ground truth. 
\begin{table*}[h]
	\centering
	\caption{Inference of randomly generated ARX networks}\label{t1}
	\resizebox{1.5\columnwidth}{!}{
	\begin{tabular}{|P{1cm} |P{1cm}|P{1cm}|P{1.2cm}|P{1cm}|P{1cm}|P{1.2cm}|P{1cm}|P{1cm}|P{1.2cm}|}
		\hline
		&\multicolumn{3}{|c|}{30dB}&\multicolumn{3}{|c|}{10dB}&\multicolumn{3}{|c|}{-30dB} \\
		\hline
		&Prec& TPR &Success &Prec& TPR &Success&Prec& TPR &Success\\
		\hline
		GESBL    & 100\% & 99.2\% & 77\%& 100\% & 97.3\%   & 6\%  & 99.6\%& 44.5\%   & 0\\
		SBL    & 100\% & 97.8\% & 34\%& 59.5\%& 95.9\%   & 0    & 53.3\%& 94.9\%   & 0\\
		GSBL   & 100\% & 76.4\% & 0   & 76.7\%& 82.2\%   & 0    & 64.4\%& 86.9\%   & 0\\
		Kernel & 100\% & 97.3\% & 29\%& 88.4\% & 90.8\%  & 0    & 80.1\%& 53.0\%   & 0\\
		\hline
	\end{tabular}
}
\end{table*}

Table~\ref{t2} shows that the estimation error of our method is the lowest with low noise ($10dB$ and $30dB$). With high noise ($SNR=-30dB$), our method is similar to the kernel method. Note that the error of SBL and GSBL reaches unacceptable levels ($146.3$ and $47.6$) if the noise is high. 
\begin{table}[h]
	\centering
	\caption{NRMSE of randomly generated ARX networks}\label{t2}
	\begin{tabular}{|P{1.2cm} |P{1.2cm}|P{1cm}|P{1.6cm}|}
		\hline
		&  30dB  & 10dB & -30dB \\
		\hline
		GESBL           & 0.21      & 0.44 & 3.75\\
		SBL           & 0.23       & 0.60 & 146.3\\
		GSBL          & 0.64       & 0.84 & 47.9\\
		Kernel        & 0.60       & 0.84 & 3.65\\
		\hline
	\end{tabular}
\end{table}

\begin{figure}[h]
	\centering
	\includegraphics[width=50mm]{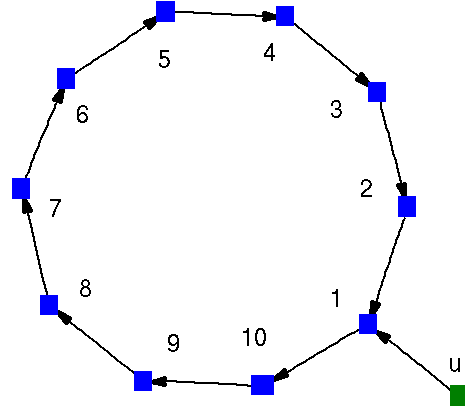}
	\caption{A ring network structure used for simulation.\label{overflow} }\label{fig}
\end{figure}

The second simulation generated 100 new random networks, this time from a fixed topology: a ring network as in Figure~\ref{fig}. There is only one input applied to node 1 and SNR was set to $20dB$. Models were generated in the same way as the first simulation. For identification, $65$ data points for each node and input were collected. This is a challenging example as it contains a feedback loop and it is very sparse. 

The inference of ring networks further highlights the significance of Prec. The ring network contains only $10$ links (of a total of $90$ possible links). Hence, high TPR is meaningless unless Prec is also high, or otherwise there is a low probability to choose true links from all of those inferred. Thus, the main task of inference in this case is to achieve both high TRP and Prec. The results are summarised in Table~\ref{t3}, which confirms this consideration: while all methods attain very high TPR ($>90\%$), only ours has high Prec ($93.2\%$). Both SBL and GSBL only have around $13\%$ chance of picking a correct link from all inferred links, meaning that these methods estimated almost all 90 possible links. The kernel method is substantially better, but still at $46\%$ Prec, the chance of getting a correct link from all other inferred links is nearly as good as flipping a coin: on average, it estimated $20.5$ links, but only $9.5$ of them are correct. Hence, only GESBL provides a useful inference as there is a high confidence that an inferred link is a correct one ($93.2\%$): on average 9.2 links are correctly inferred (out of 10) in a total of 9.9 links estimated (out of 90).  Moreover, our method  perfectly estimated ($100\%$ TPR and $100\%$ Prec) $49\%$ of all networks and had the lowest estimation error among all methods.
\begin{table}[h]
	\centering
	\caption{Inference of ring networks ($SNR=20dB$)}\label{t3}
	\begin{tabular}{|P{1cm} |P{1cm}|P{1cm}|P{1.2cm}|P{1.2cm}|}
		\hline
		&  Prec  & TPR & Success & NRMSE\\
		\hline
		GESBL    &  93.2\% & 92.4\% &49\% &1.48\\
		SBL    &  11.2\% & 100\%  &0    &9.04\\
		GSBL   &  13.2\% & 97.9\% &0    &  7.99\\
		Kernel &  46.4\% & 95.2\% &0    & 2.62\\
		\hline
	\end{tabular}
\end{table}

Results reveal that GSBL searches for a sparse network topology by setting  groups of parameters to 0, but is not able to explore element sparsity because each group of the parameter vector shares the same hyperparameter. Hence, the estimation accuracy of model parameters is poor, which in turn degrades its ability to detect topology. In contrast, SBL is sensitive to element sparsity of parameters. However, SBL is not robust to process noise so it is not able to promote sparse topologies. As SNR decreases, SBL tends to produce a fully connected network. A crucial fact is that imposing sparse topologies or parameter sparsity alone is insufficient to correctly infer networks. Basically, in terms of topology and model parameters, it seems impossible to estimate only one of them accurately without a good estimation of the other.

Unlike the above two methods, the kernel approach infers the dynamics of a system by using the power of kernels and, at the same time, imposing sparse topology. As illustrated in Tables~\ref{t1} and \ref{t3}, it considerably outperforms both SBL and GSBL. Overall, in these simulations our method exhibits the best performance since it not only explores network topology but also pursues the lowest polynomial order possible. More importantly, it is able to exclude almost all the non-existing connections (Prec near $100\%$) even if SNR is low. 

\subsection{Gene regulatory network}
The repressilator model describes the transcription and translation activities among three genes and proteins. Hill functions are used to represent dynamics of transcription while degradation and translation are described by linear terms. The model is given by~\cite{nature}: 
\begin{equation}
\begin{aligned}
x_1(k+1) &= (1-\delta_1)x_1(k)+\frac{\alpha_1}{1+x_6^{n_1}(k)}+u(k)+e_1(k)\\
x_2(k+1) &= (1-\delta_2)x_2(k)+\frac{\alpha_2}{1+x_4^{n_2}(k)}+e_2(k)\\
x_3(k+1) &= (1-\delta_3)x_3(k)+\frac{\alpha_3}{1+x_5^{n_3}(k)}+e_3(k)\\
x_4(k+1) &= (1-\delta_4)x_4(k)+\beta_1x_1(k)+e_4(k)\\
x_5(k+1) &= (1-\delta_5)x_5(k)+\beta_2x_2(k)+e_5(k)\\
x_6(k+1) &= (1-\delta_6)x_6(k)+\beta_3x_3(k)+e_6(k)
\end{aligned}
\end{equation}
where
\begin{equation}
\begin{aligned}
\delta_1&=0.3,\delta_2=0.4,\delta_3=0.5,\delta_4=0.2,\delta_5=0.4, \delta_6=0.6\\
\alpha_1&=4,\alpha_2=3,\alpha_3=5,
\beta_1=1.4,\beta_2=1.5,\beta_3=1.6\\
n_1&=1,n_2=2,n_3=2.
\end{aligned}
\end{equation}
Variables $x_1,x_2,x_3$ denote the concentration of mRNAs of three genes whereas $x_4,x_5,x_6$ represent proteins, $e(k)$ denotes i.i.d. Gaussian noise, and $u(k)$ presents the stimuli into the network (known) and was set to be a step function with amplitude $0.01$.  Parameters of the model correspond to the rate of biochemical reactions. They were chosen to produce rhythmic oscillations. The nonlinear terms in the model are Hill functions describing repressive transcriptional activity. The model was simulated with different noise variance from time indices 1 to 50.

 Assuming no prior knowledge of the network, we built a dictionary of candidate functions including linear functions and Hill functions with the Hill coefficient from 1 to 4 in both repression and activation forms. For the $i$th node, there were $9$ basis functions:
\begin{equation}
\begin{aligned}
\left[x_i, \frac{x_i}{1+x_i}, \frac{1}{1+x_i}, \frac{x_i^2}{1+x_i^2}, \frac{1}{1+x_i^2},..., \frac{x_i^4}{1+x_i^4}, \frac{1}{1+x_i^4}\right]
\end{aligned}
\end{equation}
Hence, the target vector $w$ for each node has dimension $55$. $w$ contains $6$ groups, each of which corresponds to the regulation from one gene. There are $10$ elements in each group describing the dynamics of regulations.

%Successful identification needs to not only infer the correct network topology but also indicate the type of the gene regulation dynamics by selecting the correct linear terms or Hill functions.

Table~\ref{t0} compares the inferred Boolean structure of the repressilator network. With no process noise, our method and SBL achieve perfect inference ($TPR=100\%$, $Prec=100\%$). With relatively small noise variance ($10^{-3}$), while SBL and GSB inferred all the true links, their Prec decreases to $53\%$ and $45\%$ respectively, indicating that it becomes difficult to tell which links are true in their inferred networks. In contrast, our method retains high Prec (81\%). As the noise variance increases to $10^{-1}$, Prec of SBL and GSBL further drops to $36\%$. In contrast, our method retained relatively high Prec at $76\%$: on average, it identified $6.6$ links in total, among which only $1.6$ links were wrong.
\begin{table*}[h]
	\centering
	\caption{Inference results of the Repressilator network.}\label{t0}
	\begin{center}
		\resizebox{1.5\columnwidth}{!}{
		\begin{tabular}{|c|c|c|c|c|c|c|c|c|c|}
			\hline
			&\multicolumn{3}{|c|}{No noise}&\multicolumn{3}{|c|}{$10^{-3}$ Var}&\multicolumn{3}{|c|}{$10^{-1}$ Var} \\
			\hline
			&Prec& TPR &Success &Prec& TPR &Success&Prec& TPR &Success\\
			\hline
			GESBL    & 100\% & 100\% & 100\%& 81\% & 98\%   &12\% & 76\%& 84\%   & 8\%\\
			SBL    & 100\% & 100\% &100\%& 53\% & 100\%   &0 & 36\%& 100\%   & 0\\
			GSBL   & 75\%  & 100\% & 0& 45\% & 100\%   &0 & 36\%& 100\%   & 0\\
			\hline
		\end{tabular}
	}
	\end{center}
\end{table*}

Table~\ref{t01} shows that with no process noise, the estimation errors of both our method and SBL are negligible. With noise, the estimation accuracy of our method is slightly better than SBL, while GSBL is considerably worst.
\begin{table}[h]
	\centering
	\caption{NRMSE of the Repressilator network.}\label{t01}
	\begin{center}
		\begin{tabular}{|c|c|c|c|}
			\hline
			&  No noise  & $10^{-3}$ Var & $10^{-1}$ Var\\
			\hline
			GESBL           &  1e-3 & 0.94 & 3.90\\
			SBL           &  1e-3 & 1.28 & 4.07\\
			GSBL          &  3.4    & 6.89  & 6.16\\
			\hline
		\end{tabular}
	\end{center}
\end{table}

In summary, simulation results again highlight the advantage of our approach. The prior introduced by SBL is too loose to impose group sparsity as each element of $w$ is controlled by an independent hyperparameter. In contrast, GSBL is weak in estimating the value of $w$ because elements within a group share a single hyperparameter. Our approach combines SBL and GSBL in ways that they compensate for each other. The proposed method imposes group sparsity by introducing a lower dimensional hyperparameter space (as GSBL) and also allows a sufficient degree of freedom to explore the value of $w$ (as SBL).

\subsection{Arabidopsis circadian clock model}

  In above two simulations, the ground truth models are contained in the proposed model class, which is highly unlikely for practical networks (e.g. biological networks). Next, we test our method when ground truths do not fall in the proposed model class. Many real-world networks are not ARX, but they can be well described by NARX models (e.g. circadian clocks of plants). Here, we simulated a synthetic model (Millar 10 model~\cite{Millar10}), built to describe the circadian clock of \textit{Arabidopsis}. The model is based on the real experimental data, capturing key system dynamics of the real circadian clock. In addition, it has been widely used to test different network inference methods. Millar 10 describes a circadian clock consisting of $7$ genes along with their associated proteins. In total, there are $19$ nodes in the network. The system is driven by light signals. The detailed mathematical model can be found in~\cite{Millar10}.
    
    The model simulated typical experimental conditions: for four days of light-dark cycles (data captured on LD for one day) followed by three days of constant light (data captured on LL for one day). The simulation was repeated $50$ times and data were sampled every hour. To avoid capturing transition due to  initial conditions, the simulated data for the first two days were discarded. Four series of data under LDLD (0h-44h), LDLL (24h-68h), LLLL (48h-92h) and steady state (72h-116h) were collected for inference. Typically, only mRNA concentrations are measured in high throughput biological experiments, so we assume that only those are available from the simulation. The target is to infer the network at the transcriptional (mRNA) level of $7$ clock genes.
    
    A simplified grey-box model is established to describe the Millar 10 network as follows:
    \begin{equation}
    \begin{aligned}
    \frac{dx_i(t)}{dt} = &-v_ix_i(t)\\
    &+\sum_{u=1,u\neq i}^7\sum_{j=1}^nv_{uj}^I\frac{x_u(t)}{x_u(t)+K_j}+v_{uj}^{II}\frac{K_j}{x_u(t)+K_j}\\
    \end{aligned}
    \label{semi}
    \end{equation}
    where $x_i$ denotes the $i$th gene. $v$ and $K$ are model parameters related to biochemical reaction rates. Michaelis-Menten kinetics and sigmoidal equations typically describe biochemical reactions, so our basis functions is composed of those functions: $\frac{x_u(t)}{x_u(t)+K_j}$ and $\frac{K_j}{x_u(t)+K_j}$ correspond to active and repressive transcriptional activities of gene $u$, respectively. The values of $K$ are pre-fixed and range from $0.5$ to $5$ with the increment of $0.5$. Hence, overall there are a total of $120$ basis functions for each target gene. Note that this model cannot interpret Millar 10 precisely since several nonlinearities are not present in our model class.
    
    iCheMA is a very effective method to infer biological networks. This method was shown to outperform many existing inference methods, including hierarchical Bayesian regression, LASSO, elastic net, etc (through Monte Carlo simulations on the Millar 10 model)~\cite{Infm2}. Therefore, we compare our method with this state-of-the-art method.
    
Since model~\eqref{semi} is continuous-time, we need to evaluate its derivatives. We will use the method as in iCheMA~\cite{Infm2}, based on Gaussian processes. From the estimated derivatives, the identification problem was reformulated as a linear regression: $Y = \Phi V$ where $Y$ contains derivatives of $x_i$, $\Phi$ is the dictionary matrix of Michaelis-Menten kinetics and $V$ is a vector of model parameters, $v$.
    
For a fair comparison, we use the same criteria from iCheMA to evaluate algorithm performance: the area under the receiver operating characteristic curve (AUROC) and the area under
the precision recall curve (AUPREC)~\cite{Infm1}. They are calculated based on the confidence of the inferred links. For our method, the confidence of link $j\rightarrow i$ is calculated as $P(j\rightarrow i)=\frac{\|V_j\|}{\|V\|}$ where $V_j$ contains a group of model parameters $v$ in~\eqref{semi} representing the regulation from the $j$th gene to the $i$th gene. A good inference result is indicated by high AUROC and AUPREC. 
    \begin{table*}[h]
    	\centering
    	\caption{Inference results of the circadian clock model.}\label{t06}
    	\begin{center}
    		\resizebox{1.5\columnwidth}{!}{
    		\begin{tabular}{|c|c|c|c|c|c|c|c|c|}
    			\hline
    			&\multicolumn{2}{|c|}{LDLD}&\multicolumn{2}{|c|}{LDLL}&\multicolumn{2}{|c|}{LLLL} &\multicolumn{2}{|c|}{Steady State}\\
    			\hline
    			&AUROC& AUPREC&AUROC& AUPREC &AUROC& AUPREC&AUROC&AUPREC \\
    			\hline
    			GESBL    &64.6 \% & 56.9\%&72.7 \% & 71.9\% &73.4 \% & 70.1\% &69.2\%&61.1\%   \\
    			iCheMA &66.4 \% & 62.3\%& 65.4\%& 64.2\%&69.4\% &66.8\%&64.7\%&56.4\%\\
    			\hline
    		\end{tabular}
    	}
    	\end{center}
    \end{table*}

Table~\ref{t06} indicates that our method outperforms iCheMA in almost all cases. Our method is able to provide reliable inference results where most of the true links are inferred with high confidence. In addition, our method has a considerably much higher computational efficiency than iCheMA. Indeed,  iCheMA requires more than $3$ hours to complete an inference whilst our method only demands a few minutes. This is due to the fact that iCheMA conducts a combinatorial search for basis functions whereas GESBL avoids this by imposing sparsity.

\section{Conclusion}\label{sec:Conclusion}
This paper combines SBL and GSBL to identify multivariable ARX models given measured time series data. Only limited prior knowledge of the system is needed. To infer sparse networks, the proposed model considers both topology and model complexity simultaneously. This is achieved by inducing both group and element sparse priors,  imposing both a sparse model structure and the lowest system order possible. The resulting optimisation problem is a blend of reweighed LASSO and group LASSO, which further indicates the efficiency of our approach to impose both kinds of sparsity. This framework is further extended to nonlinear systems. Simulation examples illustrate the advantages of the method.

Overall, the value of this approach is that model complexity (polynomial orders) and sparsity of the network topology are both explored at the same time. In addition, no manual tuning is required. Our method is particularly useful for NARX models, where group sparsity, in terms of sparse topology, and element sparsity, in regards to the sparse basis selection, are equally important. However, our approach may not be the most appropriate to identify other more complex models (e.g. ARMAX and dynamical structure functions (DSF)). Stability of these models is not reflected by the prior distributions in our work.    

Further developments should ideally include two aspects. The first is to obtain theoretical guarantees of the algorithm performance. Since the dictionary matrix $\Phi$ correlates with process noise due to the intrinsic property of dynamic systems, its analysis is much more complex than a pure linear regression case as in~\cite{prior1}. The second question is how to extend this framework to infer more general network models. The main obstacle here is that to estimate parametrized models, SBL normally demands the logarithm of the likelihood function to be quadratic, which does not naturally occur with general model classes. Although model parameters still hold similar sparse properties (element and group sparsity), these models are usually identified in a non-parametric way (e.g. the kernel method in~\cite{nonp}).

% or


\appendices
\section{Derivation of Type II maximization problem}\label{sec:app1}
Firstly, note that:
\begin{equation}
\begin{aligned}
&-2\log\int p(y|w)\hat{p}(w)dw \\
&-2\log\int \mathcal{N}(y|\Phi w,\lambda I)\mathcal{N}(w | \varepsilon,B) \mathcal{N}(w | 0,\Gamma)dw \\
&-2\log\int \exp(E_w)dw+ \log|\lambda I|+\log|B|+\log|\Gamma|,
\label{1}
\end{aligned}
\end{equation}
where
\begin{equation}
\begin{aligned}
E_w=-\frac{1}{2}[\lambda^{-1}\|(y-\Phi w)\|_2^2+\|w\|^{2}_{\Gamma^{-1}} +\|w-\varepsilon\|^{2}_{B^{-1}}],
\end{aligned}
\end{equation}
ignoring all the constant terms.  By completing the square:
\begin{equation}
\begin{aligned}
& -\frac{1}{2\lambda}\|(y-\Phi w)\|_2^2-\frac{1}{2}[\|w\|^{2}_{\Gamma^{-1}} +\|w-\varepsilon\|^{2}_{B^{-1}}]\\
&=-\frac{1}{2}\left[(w-\mu)^T\Sigma^{-1}(w-\mu)+E_y\right],
\label{2}
\end{aligned}
\end{equation}
where
\begin{equation}
\begin{aligned}
\Sigma&=\left[(\Gamma^{-1}+B^{-1})+\lambda^{-1}\Phi^T\Phi\right]^{-1}\\
\mu&=\Sigma(\lambda^{-1}\Phi^Ty-B^{-1}\varepsilon)\\
E_y&=\min_{w} \lambda^{-1}\|(y-\Phi w)\|_2^2+\|w\|^{2}_{\Gamma^{-1}} +\|w-\varepsilon\|^{2}_{B^{-1}}.
\end{aligned}
\end{equation}
In addition, note that:
\begin{equation}
\begin{aligned}
&-2\log\int \exp\left\{-\frac{1}{2}[(w-\mu)^T\Sigma^{-1}(w-\mu)]\right\}dw\\
&=-\log|\Sigma|\\
&=\log\left|(\Gamma^{-1}+B^{-1})+\lambda^{-1}\Phi^T\Phi\right|,
\label{3}
\end{aligned}
\end{equation}
ignoring all the constant terms.

Using~\eqref{2} and~\eqref{3}, the integral in~\eqref{1} becomes:
\begin{equation}
\begin{aligned}
&-2\log\int p(y|w)\hat{p}(w)dw \\
&=E_y + \log\left|(\Gamma^{-1}+B^{-1})+\lambda^{-1}\Phi^T\Phi\right|\\
& + \log|\lambda I|+log|B|+\log|\Gamma| \\
&=E_y + \log\left|I+(\Gamma^{-1}+B^{-1})^{-1}\lambda^{-1}\Phi^T\Phi\right|\\
&+\log|(\Gamma+B)|+\log|\lambda I|\\
&=E_y + \log\left|I+\lambda^{-1}\Phi(\Gamma^{-1}+B^{-1})^{-1}\Phi^T\right|\\
&+\log|(\Gamma+B)|+\log|\lambda I|\\
&=E_y + \log\left|\lambda I+\Phi(\Gamma^{-1}+B^{-1})^{-1}\Phi^T\right|+\log|(\Gamma+B)|\\
&=\min_{w}\lambda^{-1}\|(y-\Phi w)\|_2^2+ \|w\|^{2}_{\Gamma^{-1}} +\|w-\varepsilon\|^{2}_{B^{-1}} \\
&+ \log\left|\lambda I+\Phi(\Gamma^{-1}+B^{-1})^{-1}\Phi^T\right|+\log|(\Gamma+B)|.
\label{4}
\end{aligned}
\end{equation}
As a result, we get:
\begin{equation}
\begin{aligned}
&\mathcal{L}: \min_{\beta , \gamma,\lambda,w}\lambda^{-1}\|(y-\Phi w)\|_2^2+ \|w\|^{2}_{\Gamma^{-1}} +\|w-\varepsilon\|^{2}_{B^{-1}}\\
&+\log|B+\Gamma|+\log|\lambda I+\Phi(\Gamma^{-1}+B^{-1})^{-1}\Phi^T|\\
&\textrm{subject to:}\\
& \quad \quad \quad \quad \quad \quad  \beta\geq0 , \gamma\geq0, \lambda \geq0.\\
%&\textrm{Or:}\\
%&\mathcal{L}_2:\\
%&\min_{\beta , \gamma,\lambda}y^T\left[\lambda I+\Phi(\Gamma^{-1}+B^{-1})^{-1}\Phi^T\right]^{-1}y\\
%&+\log|B+\Gamma|+\log|\lambda I+\Phi(\Gamma^{-1}+B^{-1})^{-1}\Phi^T|\\
%&\textrm{Subject to:}\\
%& \quad \quad \quad \quad \quad \quad  \beta\geq0 , \gamma\geq0, \lambda \geq0.
\end{aligned}
\end{equation}

\section{Convexity of DCP problem}\label{sec:app2}
To see functions $u(w,\lambda,\beta,\gamma)$ and $v(\lambda,\beta,\gamma)$ are jointly convex functions, we need to prove each term is jointly convex. To check the convexity of $u(w,\lambda,\beta,\gamma)$, we consider the epigraph of its two terms defined as $epi f = \left\{(x,t)|x\in dom f, f(x)\leq t \right\}$~\cite{convexopt}. It is known that:
\begin{equation}
\begin{aligned}
&\lambda I\succ0,\quad \lambda^{-1}\|(y-\Phi w)\|_2^2<t\\
&\textrm{equivalent to}\\
&\left[\begin{array}{cc}\lambda I&y-\Phi w\\(y-\Phi w)^T&t\end{array}\right]\succ0,
\end{aligned}
\end{equation}
so the term $\lambda^{-1}\|(y-\Phi w)\|_2^2$ is jointly convex as is same with $ \|w\|^{2}_{\Gamma^{-1}}$ and $\|w-\varepsilon\|^{2}_{B^{-1}}$ since their epigraphs are convex sets described by LMIs~\cite{convexopt}.

For the function $v(\lambda,\beta,\gamma)$, firstly note that $-\log|\cdot|$ is a convex function in $S^{+}$. Since $B+\Gamma$ is an affine function of $\Gamma$ and $B$, $-\log|B+\Gamma|$ is jointly convex with respect to $\Gamma$ and $B$. The second term of the function $v$ seems more complex. To prove its convexity, we first consider the following lemma. A similar lemma with lower dimension of function domain can be found in~\cite{convexopt}:

\begin{lem}
	Suppose a function $f(x)=h(g_1(x),...,g_k(x))$ where $h(z_1,...,z_k):R^k\rightarrow R$ and $g_r:R^n\rightarrow R$. Then $f$ is concave if $h$ is concave and nondecreasing in each argument and $g_r$ is concave.
\end{lem}

The proof of this lemma is straightforward by checking the Hessian matrix:
\begin{equation}
\begin{aligned}
\nabla^2f = J^T_g\nabla^2hJ_g+\sum_{l=1}^k \frac{\partial h}{\partial z_l}\big|_{g_l}\nabla^2g_l,
\end{aligned}
\end{equation}
where $J_g$ denotes the Jacobian matrix of the function $g=[g_1,...,g_k]^T$.

Let $h(x^1,x)=\log|x^1I+\Phi diag(x)\Phi^T|$ and $g^1(\beta,\gamma,\lambda)=\lambda$, $g_{rj}(\beta,\gamma,\lambda)=\frac{\gamma_r\beta_{rj}}{\gamma_r+\beta_{rj}}$ with $x\in R^{(p+m)k}$, $r\in[1,p+m]$ and $j\in[1,k]$, then $h(g^1,g_{11},g_{12},...,g_{(p+m)k})=\log|\lambda I+\Phi(\Gamma^{-1}+B^{-1})^{-1}\Phi^T|$. Obviously, $h(\cdot)$ is jointly concave with respect to $x^1$ and $x$.

Since the Hessian matrix of $g^1$ is $0$, we only need to check the gradient of $h(\cdot)$ with respect to $x$:
\begin{equation}
\begin{aligned}
%\frac{\partial h}{\partial x^1}&=trace\left[(x^1I+Adiag(x)A^T)^{-1}\right]\\
%&=trace(\Delta^1)\\
\frac{\partial h}{\partial x_r}&=trace\left[\Phi^T(x^1I+\Phi diag(x)\Phi^T)^{-1}\Phi diag(e_r)\right]\\
&=\Delta_{rr},
\end{aligned}
\end{equation}
where $\Delta=\Phi^T(x^1I+\Phi diag(x)\Phi^T)^{-1}\Phi$ and $e_r$ is a vector with its $r$th element $1$ and all the others $0$.

Since matrix $\Delta$ is at least semi-positive definite, its diagonal elements must be non-negative. As a result, $h(x)$ is nondecreasing in each argument. We finally calculate the Hessian matrix, $H$ of $g_{rj}(\beta,\gamma,\lambda)$. Note that matrix entries $H_{qq}=\frac{\partial^2 g_{rj}}{\partial \beta_{rj}^2}$, $H_{ll}=\frac{\partial^2 g_{rj}}{\partial \gamma_r^2}$ and $H_{ql}=H_{lq}=\frac{\partial^2 g_{rj}}{\partial \beta_{rj}\partial\gamma_r}$ with all the other entries 0 where $q=(r-1)k+j$ and $l=(p+m)k+r$. It is always possible to find a permutation matrix, $P$ such that:
\begin{equation}
\begin{aligned}
P^THP=\left[\begin{array}{cc}\hat{H}&\textbf{0}\\ \textbf{0}&\textbf{0}\end{array}\right],
\end{aligned}
\end{equation}
where
\begin{equation}
\begin{aligned}
\hat{H}=\left[\begin{array}{cc}-\frac{2\gamma_r^2(\gamma_r+\beta_{rj})}{(\gamma_r+\beta_{rj})^4}&\frac{2\gamma_r\beta_{rj}(\gamma_r+\beta_{rj})}{(\gamma_r+\beta_{rj})^4}\\ \star &-\frac{2\beta_{rj}^2(\gamma_r+\beta_{rj})}{(\gamma_r+\beta_{rj})^4}\end{array}\right].
\end{aligned}
\end{equation}
Obviously, matrix $\hat{H}$ is semi-negative definite so that $H$ is also semi-negative definite, which indicates function $g_{rj}(\beta,\gamma,\lambda)$ is concave. For $g^1$, it is an affine function. According to the lemma above, $\log|\lambda I+\Phi(\Gamma^{-1}+B^{-1})^{-1}\Phi^T|$ is jointly concave with respect to $\lambda$, $\gamma$ and $\beta$.

%\section*{Acknowledgment}

\ifCLASSOPTIONcaptionsoff
  \newpage
\fi



%\bibliographystyle{IEEEtran}
%\begin{thebibliography}{1}
%\end{thebibliography}
\bibliographystyle{IEEEtran}
\bibliography{IEEEabrv,IEEEexample}
%\end{IEEEbiography}
%% insert where needed to balance the two columns on the last page with
%%\newpage
%% You can push biographies down or up by placing
%% a \vfill before or after them. The appropriate
%% use of \vfill depends on what kind of text is
%% on the last page and whether or not the columns
%% are being equalized.
%
%%\vfill
%
%% Can be used to pull up biographies so that the bottom of the last one
%% is flush with the other column.
%%\enlargethispage{-5in}
%
%

\end{document}